\documentclass[preprint,aps,showkeys,showpacs]{revtex4}
\usepackage{graphicx}
\usepackage{dcolumn}
\usepackage{bm}

\preprint{}
\def\para{\par\noindent}
\def\sqr#1#2{{\vcenter{\vbox{\hrule height.#2pt
        \hbox{\vrule width.#2pt height#1pt \kern#1pt
          \vrule width.#2pt}
        \hrule height.#2pt}}}}

\newcount\notenumber

\def\note{\advance\notenumber by 1
\footnote{$^{\the\notenumber}$}} \baselineskip 20pt
\begin{document}
\title{Persistence in Financial Markets}
\author{S.Jain and P.Buckley \\ Information Engineering,\\ The Neural
Computing Research Group,\\ School of Engineering and Applied
Science,\\ Aston University, \\ Birmingham B4 7ET, \\ U.K.}

\begin{abstract}

 Persistence is studied in a financial context by mapping the time
evolution of the values of the shares quoted on the London Financial
Times Stock Exchange 100 index (FTSE 100) onto Ising spins.
 By following the time dependence
of the spins, we find evidence for power law decay of the proportion
of shares that remain either above or below their \lq starting\rq\
values. As a result, we estimate a persistence exponent for the
underlying financial market to be $\theta_f\sim 0.5$.
\end{abstract}

\maketitle
\section{Introduction}
\para In its most general form, persistence is concerned
 with the fraction of space
which persists in its initial state up to some later time. The
problem has been extensively studied in recent years for pure spin
systems at both zero [1-4] and non-zero [5] temperatures.
\para For example, in the non-equilibrium dynamics of spin systems
 at zero-temperature,
 the system is prepared in a random state at $t=0$ and the fraction of
spins, $P(t)$, that persist in the same state as at $t=0$ up to some
later time $t$ is studied. For the pure ferromagnetic
two-dimensional Ising model the persistence probability has been
found to decay algebraically [1-4]
$$ P(t)\sim t^{-\theta},\eqno(1)$$
where $\theta\sim 0.22$ is the non-trivial persistence exponent
[1-3].
\para The value of $\theta$ depends on both the spin [6] and
spatial [3] dimensionalities; see Ray [7] for a recent review.
\para At non-zero temperatures [5], consideration of the global order
parameter leads to a value of $\theta_{global}\sim 0.5 $ for the
two-dimensional Ising model.
\para Very recently, disordered systems [8-10] have also been studied
and have been found to exhibit different persistence behaviour to
that of pure systems.
\para Persistence has also been studied in a wide range of experimental
systems and the value of $\theta$ ranges from $0.19$ to $1.02$
[11-13].
 Much of the recent
theoretical effort has gone into obtaining the numerical value of
$\theta$ for different models.

\para In this work we present the first estimate for a persistence
exponent extracted from {\it financial} data.
\para Long-range correlations in persistent and
anti-persistent random walks were first discussed by Mandelbrot
[14]. Zhang [15] has presented empirical evidence to support that
daily returns in composite indices are not completely randomized.
Here, on the other hand, we study the behaviour of the constituent
stock prices.

\section{Financial Markets}
\para A financial market is an example of a complex
many-body system exhibiting many of the characteristics found in
model systems studied in statistical physics.

There is an element of both co-operation and \lq frustration\rq\
[16] in the movement of share values. For example, share values of
companies in the same sector tend to move in the same direction
(either up or down) when subjected to identical external events. A
typical case here would be the movement in the value of shares in
oil companies on the news of over/under production. On the other
hand, there are also companies whose share values move in opposite
directions given the same event. Here, typical examples would be the
reactions in the share values of companies from the retail and
banking sectors on the news of an increase/decrease in interest
rates.

\para In this work we make no assumptions about any underlying model
systems. Rather, we treat the historical share values of the
companies over time as the outcomes of some \lq experiment\rq\.
\para The financial market we
study is the set of companies quoted on the London Financial Times
Stock Exchange 100 (FTSE 100) share index.
\section{Method}
\para The data used in this study was obtained from Datastream [17], a
database of financial information, and refers to the end of day
prices over the randomly chosen ten year
 period from 24 February 1995 to 1 February 2005. The data were mapped
onto Ising spins using the procedure outlined below.
\para  The \lq base\rq\ share price, $(P^b_i(t=0), i=1\dots , 100)$, of
each of the companies appearing in FTSE 100 at the end of trading on
24 February 1995 was noted to 2 decimal places.
  All prices
 for the shares used in this work were taken to be the closing values
at the end of trading. At $t=1$ (the end of trading on the next day)
the share price of each company,
 $P_i(t=1), i=1\dots , 100$, was compared with the corresponding base price.

\para We allocate a value $S_i(t=0)=+1$ if $P_i(t=1)\ge P^b_i(t=0)$ and
a value of $S_i(t=0)=-1$ if $P_i(t=1)<P^b_i(t=0)$.
\begin{table}
\centering
\begin{tabular}{|c|c|c|c|}
  \hline
  {Date} & {t} & {Closing price} & {$S_i(t)$} \\ \hline
   24 February 1995 & 0 & 166.00 & Base price \\ \hline
   27 February 1995 & 1 & 166.50 & +1 \\ \hline
   28 February 1995 & 2 & 172.75 & +1 \\ \hline
   1 March 1995 & 3 & 167.00 & +1 \\ \hline
   2 March 1995 & 4 & 161.75 & -1 \\
  \hline
\end{tabular}
\caption{A mapping onto Ising spins of the end of day closing share
prices for a typical company quoted on the FTSE 100.}
\label{spinvalues}
\end{table}
Table 1 gives a typical example of the mapping. Note
that the value of the spin is determined with reference to the base
price and not the previous closing price. Furthermore, in this work
we disregard all fluctuations which may have taken place during the
day and simply use the end of closing prices. In the example
discussed in Table 1, as the spin has \lq flipped\rq\
when $t=4$, the value of $S_i(t\ge 4)=-1$, {\it irrespective} of
subsequent closing prices.
\para The values $\{S_i(t=0), i=1,\dots 100\}$ form the initial
configuration for our \lq spin\rq\ system. All of the subsequent
 10 years worth of data was converted into possible values of Ising
spins, $S_i(t)$, using the share values at $t=0$ as the base. As a
result, we are able to use
 $S_i(t)$ to track the value of the underlying share relative to its
base price. It is worth noting that companies have to satisfy
certain qualification criteria
 before they are included in the FTSE 100 [18]. As a result, in
practice, a given company's presence in the FTSE 100 can fluctuate
from year to year.
 In our analysis we restricted ourselves to the core set of
 companies remaining in the FTSE 100 throughout the time period under
consideration.

\para Hence, the {\it first time} $S_i(t)\ne S_i(t=0)$ corresponds to
the underlying share value either
 going above $(S_i(t)=+1)$ or below $(S_i(t)=-1)$ the base price also
for the first time. This gives us a direct analogy with the
persistence
 problem that has been extensively studied in spin systems.
 \para At each time step, we count the number of spins that still
persist in their initial $(t=0)$ state by evaluating [19]
 $$n_i(t)=(S_i(t)S_i(0)+1)/2.\eqno(2)$$
 Initially, $n_i(0)=1$ for all $i$. It changes to zero when a spin
flips (that is, the underlying share price goes above/below the base
price) for the first time. Note that
 once $n_i(t) =0$, it remains so for all subsequent calculations.
\para The total number, $n(t)$,
of spins which have never flipped until time $t$ is then given by
$$ n(t)=\sum_{i}n_i(t).\eqno(3)$$
A key quantity of interest is $R(t)$, the density of non-flipping
spins [1]
$$R(t)=n(t)/N,\eqno(4)$$
where $N$ is the number of companies monitored (note that $N$ is not
necessarily $100$ for the reasons outlined earlier). The actual
values of $N$ used are stated below. However, it's worth noting
that, in principle, we are dealing with a model system where the
spins are interacting with all other spins. As a result, we do not
believe that our system of spins is too small.
\section{Results}
\para  We now discuss our results.
 In this initial study, three different time periods were considered:
 \begin{description}
 \item[a)] 24 February 1995 to 1 February 2005 ($N=79$)
 \item[b)] 9 January 1996 to 28 December 2000 ($N=79$)
 \item[c)] 3 January 2000 to 3 January 2005 ($N=92$)
 \end{description}
 \para These sets were selected at random. For each time period, the
 initial and subsequent spin configurations
were generated as outlined above and the resulting data analysed for
persistence behaviour.
\begin{figure}
\includegraphics{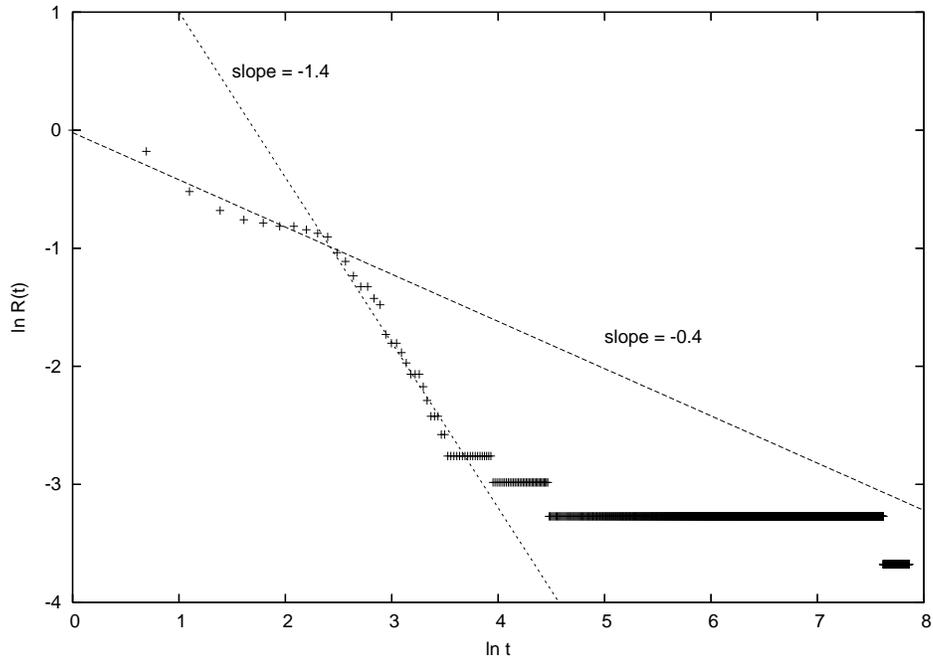}
\caption{A plot of $\ln R(t)$
 against $\ln t$ for the data commencing 24 February 1995. The two
straight lines are guides to the eye and have slopes
 $-0.4$ and $-1.4$ as indicated.}
\label{fig1}
\end{figure}
\para In Fig. 1 we show a log-log plot of the density of
non-flipping spins against time for the first set of data commencing
24 February 1995. There is evidence for an initial power-law decay
(slope = -0.4), leading to a faster subsequent decay with slope =
-1.4.
\para A key feature of Fig. 1 is that nearly all of the
spins have flipped after $\ln t\approx 4$. This corresponds to
approximately $30-50$ days of trading on the markets.

 Note that we
are not distinguishing between those shares that remain above or
below their base values. There are also a few shares (in this case
just 2) that still persist in their initial state over the entire
observation period of 10 years.
\begin{figure}
\includegraphics{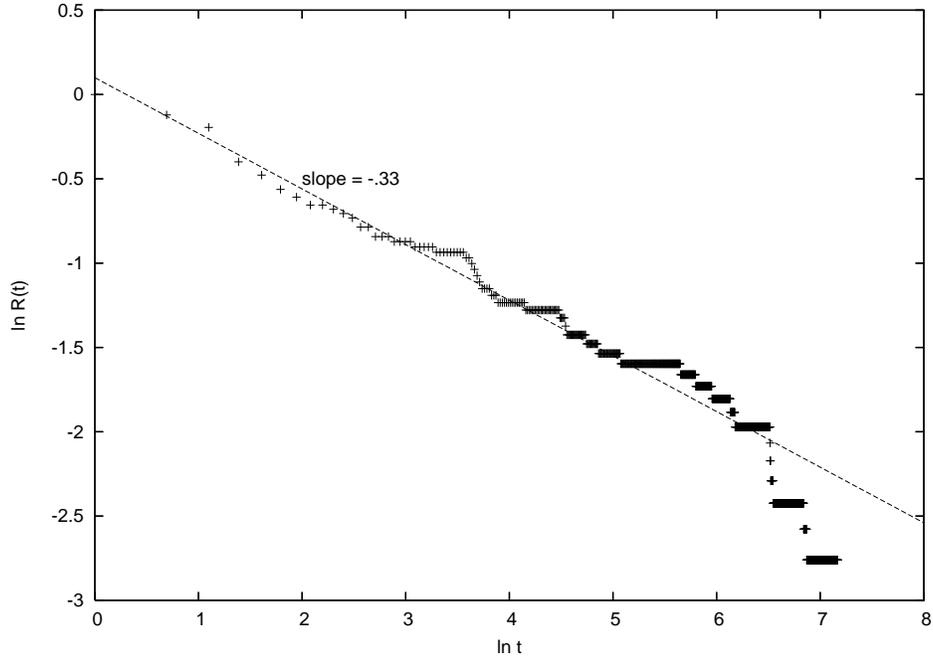}
\caption{A log-log plot of the data commencing 9 January 1996. The
straight line (guide to the eye) has a slope of $-0.33$.}
\label{fig2}
\end{figure}
To investigate the problem further, the same data was partitioned
into essentially two 5-year blocks as outlined above. The analysis
was repeated on each of the two sets of data. In Fig. 2 we
plot $\ln R(t)$ against $\ln t$ for the data commencing 9 January
1996. Note that for this plot the base prices are determined by
close of trading on 9 January 1996. Once again, there is clear
evidence for a power-law decay. This time, however, the slope of the
linear fit is $-0.33$.
\begin{figure}
\includegraphics{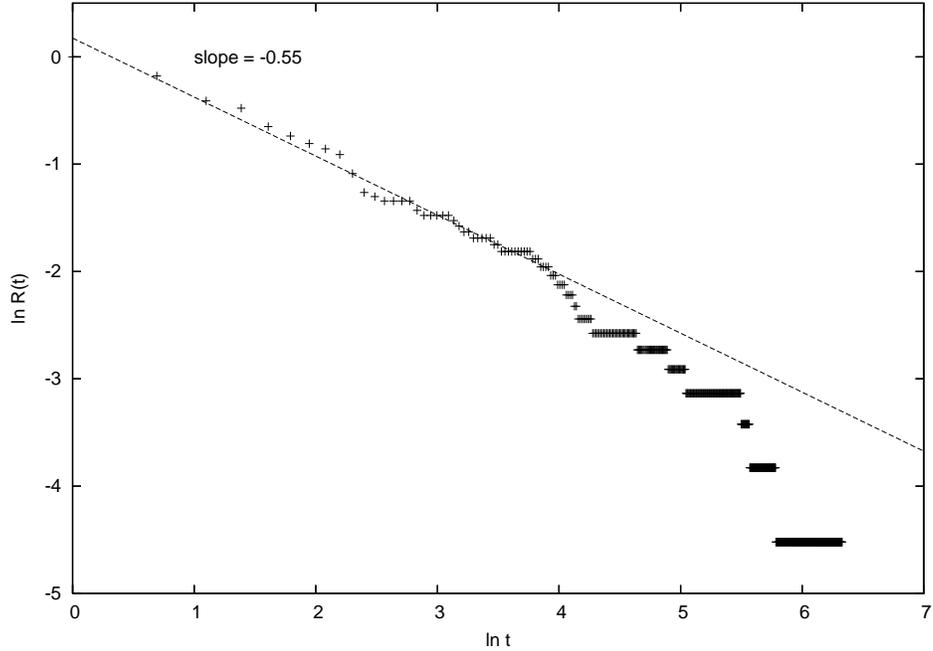}
\caption{A log-log plot of the data commencing 3 January 2000. Once
again, the straight line (slope $-0.55$) is a guide to the
 eye.}
\label{fig3}
\end{figure}
\para Finally Fig. 3 show a similar plot for the data commencing 3 January
2000. once again, we have evidence for initial power-law decay
(slope = $-0.55$) and most of the spins appear to have flipped after
$\ln t\approx 4$.

 Although, as expected, there is noise in the data, we see that
in all three cases we have clear evidence of an initial power-law
decay. It\rq s\ also clear from the plots that nearly all of the
spins have flipped after a fairly short period of time,
corresponding to approximately
 $30-50$
days of trading on the markets. However, there are a handful of
spins which do not flip over the entire time period under
consideration.

Furthermore, there appears to be a cross-over to a faster power-law
decay for longer times. This cross-over is only really evident in
Fig. 1 as there is too much noise in the other 2 sets of data. The
initial decay in Fig. 1 has a slope of $-0.4$. The faster decay is
indicated by the straight line which has a slope of $-1.4$. The
initial power law decays are indicated by the straight lines in
Figures 2 and 3. The cross-over could be a signature of the market
reacting to external events such as significant interest rate
variations or political news.

 From the linear fits,
we can extract a value of $\theta_f$ ranging from $0.33$ to $0.55$.
As a consequence, we estimate the persistence exponent for the
financial data to be $\sim 0.5$. We believe this to be the first
estimate of a persistence exponent from financial data.
\para Our value for $\theta_f$ is not inconsistent with
the value obtained from computer simulations of the $2D$ Ising model
at a non-zero temperature [5]. This is an intriguing result as we
have made no assumptions whatsover about the underlying model which
gives rise to the financial data.
 Of course, in our analysis, the value of each $S_i(t)$ incorporates the
overall performance of the shares of the underlying company relative
to the base value.
\section{Conclusion}
\para To conclude, we have used a novel mapping to map the share values
quoted on the London Financial Times Stock Exchange 100 share index
onto Ising spins. As a result, we extracted a value of $\sim 0.5$
for the persistence exponent. This should be regarded as an initial
estimate and further work is required to confirm the value. It
should be noted that, out of necessity, we worked with end of day
closing prices. Ideally, it would be better to use tick-data. It\rq
s remarkable that our value is not inconsistent with the value of
the persistence exponent obtained for the $2D$-Ising model at
non-zero temperature. This observation justifies further
investigation.

\section*{References}
\begin{description}
\item {[1]} B. Derrida, A. J. Bray and C. Godreche, J. Phys. A:
Math Gen {\bf 27},
 L357 (1994).
\item {[2]} A. J. Bray, B. Derrida and C. Godreche, Europhys. Lett.
{\bf 27},
 177 (1994).
\item {[3]} D. Stauffer J. Phys. A: Math Gen {\bf 27}, 5029 (1994).
\item {[4]} B. Derrida, V. Hakim and V. Pasquier, Phys. Rev. Lett.
{\bf 75},
 751 (1995); J. Stat. Phys. {\bf 85}, 763 (1996).
 \item {[5]} S. N. Majumdar, A. J. Bray, S. J. Cornell, C. Sire,  Phys.
Rev. Lett. {\bf 77}, 3704 (1996).
 \item {[6]} B. Derrida, P. M. C. de Oliveira and D. Stauffer, Physica
{\bf A}224, 604 (1996).
 \item {[7]} P. Ray, Phase Transitions {\bf 77} (5-7), 563 (2004).
\item {[8]} S. Jain, Phys. Rev. E{\bf 59}, R2493 (1999).
\item {[9]} S. Jain, Phys. Rev. E{\bf 60}, R2445 (1999).
\item {[10]} P. Sen and S. Dasgupta, J. Phys. A: Math Gen {\bf 37},
11949 (2004)
\item {[11]} B. Yurke, A. N. Pargellis, S. N. Majumdar
and C. Sire, Phys. Rev. E{\bf 56}, R40 (1997).
\item {[12]} W. Y.
Tam, R. Zeitak, K. Y. Szeto and J. Stavans, Phys. Rev. Lett. {\bf
78}, 1588 (1997).
\item {[13]} M. Marcos-Martin, D. Beysens, J-P
Bouchaud, C. Godreche and I. Yekutieli, Physica {\bf D}214, 396
(1995).
\item {[14]} B. Mandelbrot, International Economic Review {\bf 10},
82 (1969).
\item {[15]} Yi-C. Zhang, Physica {\bf A}269, 30 (1999).
\item {[16]} G. Toulouse, Comm. Phys. {\bf 2}, 115 (1977).
\item {[17]} Datastream [Finance, accounting and economics data
online] (2005).
\item {[18]} Fact File, London Stock Exchange,
London, England (1997).
\item {[19]} S. Jain and H. Flynn, J. Phys.
A: Math Gen {\bf 33}, 8383 (2000).
\end{description}
\end{document}